\documentclass[twocolumn,aps,prl,showpacs,superscriptaddress,amssymb]{revtex4}
\usepackage{graphicx,epsf,epsfig}
\usepackage{dcolumn}
\usepackage{bm}
\usepackage{color}
\usepackage{ulem}
\usepackage{subfigure}

\begin{document}

\title{Role of Many-particle excitations in Coulomb Blockaded Transport}

\author{B. Muralidharan}
\affiliation{School of Electrical and Computer Engineering and the Network for Computational Nanotechnology, Purdue University, West Lafayette IN 47906 USA}

\affiliation{Department of Mechanical Engineering, Massachusetts Institute of Technology, Cambridge MA02139 USA}
\author{L. Siddiqui}
\affiliation{School of Electrical and Computer Engineering and the Network for Computational Nanotechnology, Purdue University, West Lafayette IN 47906 USA}
\author{A. W. Ghosh}
\affiliation{Dept. of Electrical and Computer Engineering, Univ. of Virginia, Charlottesville VA 22904 USA}
\begin{abstract}
We discuss the role of electron-electron and electron-phonon correlations in current flow
in the Coulomb Blockade regime, focusing specifically on nontrivial signatures arising 
from the break-down of mean-field theory. By solving transport equations directly in Fock
space, we show that electron-electron interactions manifest as gateable excitations 
experimentally observed in the current-voltage characteristic. While these excitations 
might merge into an incoherent sum that allows occasional simplifications, a clear
separation of excitations into slow `traps' and fast `channels' can lead to further 
novelties such as negative differential resistance, hysteresis and random telegraph 
signals. Analogous novelties for electron-phonon correlation include the breakdown of 
commonly anticipated Stokes-antiStokes intensities, and an 
anomalous decrease in phonon population upon heating due to reabsorption of emitted 
phonons. 
\end{abstract}

\maketitle

\section{I. Introduction}
The experimental study of electron flow through nanostructures has been a 
dynamic field of activity, with an eye on extending and complementing present day 
transistor technologies, as well as generating entirely new applications. Nanoscale 
electronic transport spans a broad range of natural and artificially fabricated 
nano-structures, from carbon nanotubes, graphene nanoribbons and silicon nanowires 
\cite{lund} to spintronics and organic molecular electronics \cite{rreed,tour}. In 
particular, there has been enormous interest in quantum dot structures for exploring 
novel transport phenomena and device applications beyond the transistor switching 
paradigm, such as the exploration of double quantum dot structures \cite{kou} for 
spin-based qubit manipulation and detection \cite{kow}. Electron transport through 
natural and artificial molecules forms a key research topic, especially for switching, 
sensing \cite{ucla} and quantum computation based applications \cite{divenc}. 

In typical transport simulations, for instance in the widely implemented non-equilibrium
Green's function (NEGF) formalism, it is common to include the effect of electron-electron
or electron-vibronic interactions approximately through an effective one-electron potential
or self-consistent field (SCF) 
that needs to be computed self-consistently. There are many exceptions, however, where such 
an approximation may break down, especially when interaction energies dominate other energy 
scales of interest such as the level broadening and the device temperature. One such 
regime, well known as Coulomb Blockade (CB) \cite{rlikharev}, occurs when the device or 
channel capacitance is low enough that an active electron inside the channel can prevent 
a subsequent one from entering. Such a single particle quantization of charge 
transfer is frequently observed in chemical reactions \cite{atkins}, but is a relative 
newcomer in electronic transport measurements \cite{rlikharev}. The sequential addition
of electrons in integer amounts disallows mean-field treatments which tend to smear out 
charges and interactions by treating all electrons on an equal footing. In contrast, 
solutions involve products of electronic occupancy and atomic displacement operators including 
an exclusion principle term that requires keeping track of every possible electronic or 
vibrational configuration through the employment of the many-particle Hilbert (Fock) space. 
Under these "strong correlation" conditions, energy "levels" must 
be calculated not through a simple band theory or an effective potential, but as differences 
between {\it{total energies}} of the neutral and the cationic/anionic/vibronically excited 
species. This is extremely difficult since it requires enumeration of all many-particle 
configurations (2$^N$ of them, for N basis sets involving electronic and phononic coordinates!). 
However, such a complexity is necessary, as exclusion in Fock space creates a rich spectrum of 
excitations as well as universal scaling rules for the current plateau heights that are hard to capture 
{\it{a-priori}} using a modified one-electron potential \cite{tddft}. The alternate 
Fock space viewpoint (in general, a many-body density matrix theory) has been somewhat 
restricted to describe quantum dot transport \cite{rlikharev,rralph} and has been relatively
unexploited in molecular electronics. The focus of this article is to illustrate the Fock 
space view-point of transport, and its experimental ramifications as far as observable
signatures of many-particle excitations go. 

Our recent work in the area of molecular transport \cite{rbhasko1,rbhasko2,rbhasko3,rowen}, 
as well as compelling recent experiments triggered towards spin-based quantum computation 
\cite{kow,pet,cm,kw,now}, both argue for increased activity in this area. This article 
focusses on the role of both electron-electron and electron phonon correlations in 
non-equilibrium transport. The paper is organized into three broad sections. In the first 
section we introduce the Fock space viewpoint. The second section runs through the 
formalism of Fock-space transport. Section three discusses Coulomb Blockade signatures
created by electron-electron interactions. We elaborate on the non-trivial role of electronic 
excitations \cite{rbhasko2} in the interpretation of frequently observed I-V characteristics 
\cite{jpark,zhitenev}. It is further shown that electronic excitations can also result in intrinsic 
asymmetries within the channel that can provide an elegant approach to understanding negative 
differential resistance (NDR), hysteresis effects \cite{rbhasko3}, and random telegraph noise
\cite{ghoshieee}. The fourth section discusses the additional subtleties imposed by strong 
electron-phonon intereactions. We show that the inclusion of Fock space excitations within the 
electron-phonon manifold not only explains anomalous scaling of phonon conductance side bands 
\cite{lutfe,leroy}, but also predicts an anamolous temperature distribution of the phonon 
population. 
 
\section{II. Theoretical Background}
\begin{figure}

\centerline{\epsfig{figure=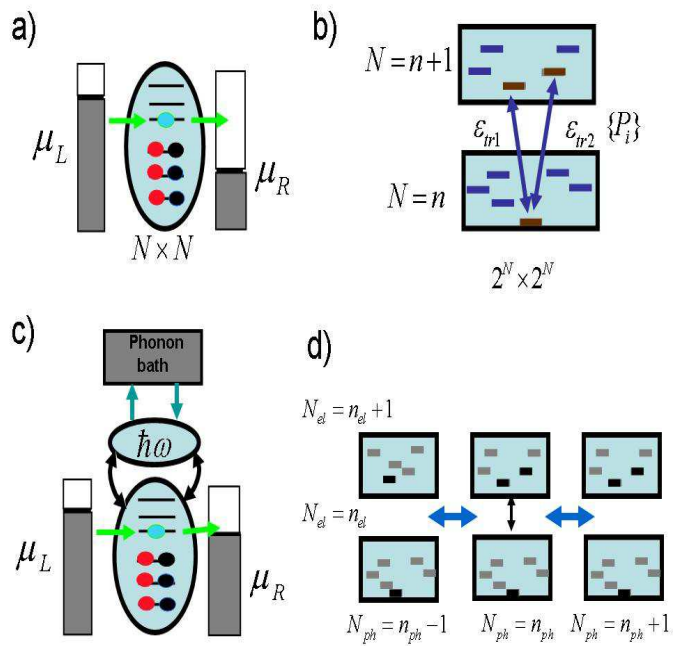,height=3.5in,width=3.5in}}
\caption{Introducing Fock Space transport. a) A schematic of electronic transport in the one-particle picture. The channel is coupled to two electrodes which add 
and remove electrons to and from the channel, resulting in current flow. Given a set of $N$ basis functions the transport problem computationally scales as 
$N \times N$. b) In the Fock space picture such addition and removal processes can be understood 
in terms of transitions between states that differ by a single electron number. In such a scheme one needs to keep track of all 
possible configurations in the channel, thus resulting in an exponential $2^N \times 2^N$ scaling of the transport problem. c) In case of strong electron-phonon interactions, 
the phonon system is coupled to a phonon bath which is maintained at equilibrium. d) The inclusion of phonons calls for using copies of the electronic Fock space, each corresponding to 
a phonon number. Electronic transport results in vertical transitions while phonon coupling results in horizontal ones.} 
\label{fig_1}
\end{figure}

The Fock space approach to Coulomb Blockaded transport was proposed originally by 
Beenakker \cite{rralph} in order to explain the CB conductance peak spacings and heights 
observed in semiconducting quantum dots. Figure~\ref{fig_1} explains the difference between 
the one-particle and many-particle Fock space pictures. A set of $N$ single particle energy 
levels generates $2^N$ Fock states corresponding to emptying or filling each of these level
with one electron. In the one-electron picture, transport involves the addition and removal 
of electrons between a set of channel levels and two macroscopic electrodes (Fig.~\ref{fig_1}(a)).
The levels themselves are computed by solving the one-electron Schr\"odinger equation, including
electronic interactions approximately through a mean-field potential that modifies these levels
dynamically. In the Fock space approach, however, the addition and removal of electrons 
leads to a transition between two entirely different multielectron configurations, in this
case, configurations that differ by a single electron (Fig.~\ref{fig_1}(b)). In other words,
electronic transport processes show up as {\it{vertical}} transitions between total energies
of electronic Fock states differing by a single electron. The situation changes a bit when 
transport involves the emission or absorption of phonons. In the one-particle picture, an
electron jumps between two one-electron levels differing by the phonon energy (Fig.~\ref{fig_1}(c)),
the phonon system itself driven towards equilibrium through a separate coupling to a
thermal bath. In the Fock space approach Fig.~\ref{fig_1}(d)), phonon-assisted transport shows 
up as {\it{horizontal}} transitions between multiple copies of the electronic Fock space that 
each correspond to a different phonon number. 

In the following, we will progressively build complexity into our Fock space approach and
attempt to identify the experimental ramifications of doing so. 

\subsection{A. Single level system, SCF analysis}
\begin{figure}[ht]
\centerline{\epsfig{figure=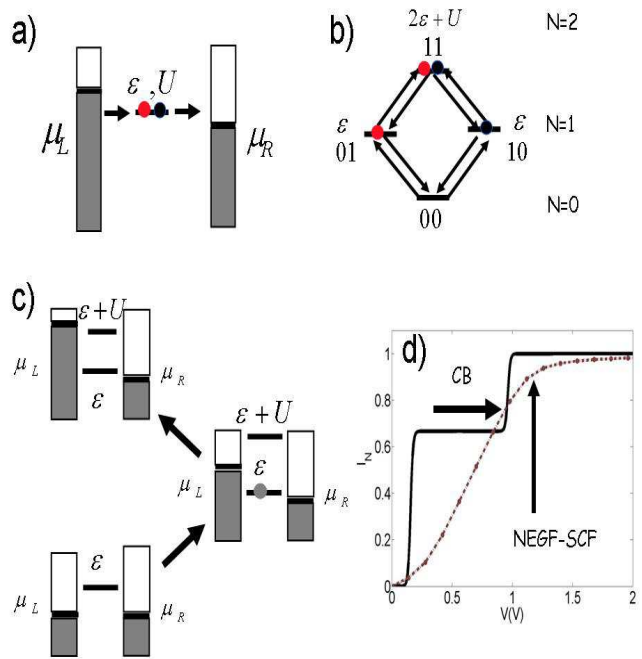,height=3.5in,width=3.5in}}
\caption{Fock Space transport through a singly degenerate energy level. a) The singly degenerate energy level can carry two electrons, one of each type. b) The Fock space thus 
comprises of $4$ states. Transitions between these result in c) two transport channels $\epsilon$ and $\epsilon + U$. Successive access of these transport channels 
results in two current jumps at different bias points separated by a plateau. A NEGF-SCF calculation is shown dotted for comparision. } 
\label{fig_2}
\end{figure}

Let us start with the smallest interacting system, namely, a real or artifical molecule
with a single energy level capable of accomodating two spins (Fig.~\ref{fig_2}a). 
The onsite energy of the level is $\epsilon_{0}$ and the Coulomb charging energy $U$.
The level is coupled to contacts which are held separately at thermal equilibrium at
their respective bias-separated electrochemical potentials $\mu_{L,R}$ ($L$: left, $R$: right). 
A starting point for our analysis is the Hubbard Hamiltonian for the molecule
\begin{equation}
\hat{H} = \epsilon\hat{n} + U\hat{n}_{\uparrow}\hat{n}_{\downarrow}
\end{equation}
where the operators $\hat{n}_{\uparrow,\downarrow}$ have eigenvalues 0 and 1, while $
\hat{n} = \hat{n}_\uparrow + \hat{n}_\downarrow$. 
Exact diagonalizing this Hamiltonian leads to a Fock space consisting of
four many-electron states (Figure~\ref{fig_2}(b)), 
an empty zero electron state $|00\rangle$ with energy $0$, two one electron states $|01\rangle$ and $|10\rangle$ 
corresponding to an up or a down spin with energy $\epsilon$, and a doubly occupied up-down spin electron state $|11\rangle$ with energy $2\epsilon_0 + U$. Equilibrium occupancies of these many-electron states are given by the Boltzmann distribution $P_N = e^{-(E_N-\mu N)/k_BT}/\Omega$, where $k_BT$ is the thermal energy, $\mu = E_F$ is the equi librium contact Fermi energy or electrochemical potential, and $\Omega = \sum_N e^{-(E_N-\mu N)/k_BT}$ is the grand partition function. The average electron occupancy is then 
given by $\langle N\rangle = \sum_N NP_N$.

One can bypass the many-electron Fock space treatment by employing a suitable 
self-consistent (SCF) potential acting in the one-electron subspace, modifying the energies 
accordingly. In the {\it{spin restricted}} approach that treats both spins equally, 
\begin{equation}
U_{RSCF} = \langle {\partial\hat{H}}/{\partial N}\rangle.
\end{equation} 
where $\langle\ldots\rangle$ denotes a quantum mechanical average. 
The interacting term can be written as
\begin{eqnarray} 
H_{int} &=& U\hat{n}_\uparrow\hat{n}_\downarrow \nonumber \\
 \hspace*{10pt} &=& U/2\sum_\sigma \hat{n}_\sigma \hat{n}_{\bar{\sigma}} \nonumber \\
 \hspace*{10pt} &=& (U/2)\sum_\sigma \hat{n}_\sigma (N-\hat{n}_\sigma) \nonumber \\
 \hspace*{10pt} &=& (U/2) N\sum_\sigma \hat{n}_\sigma - (U/2)\sum_\sigma \hat{n}_\sigma^2 \nonumber \\ 
 \hspace*{10pt} &=& UN(N-1)/2, 
\end{eqnarray}
where we have used the fact that $N=\sum_\sigma \hat{n}_\sigma$, and $\hat{n}_\sigma^2 = \hat{n}_\sigma$, 
since $\hat{n}$ can only take values of zero or one. The spin restricted SCF potential is then 
given by 
\begin{equation}
U_{RSCF} = \langle\partial H_{int}/\partial N\rangle
= U(\langle N\rangle-1/2)
\end{equation}
For a given electrochemical potential, one guesses the value of 
the average occupancy $\langle N\rangle$, uses it to calculate the SCF potential,
and then calculates in turn the mean-field occupancy $\langle N\rangle$ of the level 
$\tilde{\epsilon} = \epsilon + U_{SCF}$ using the 
Fermi-Dirac distribution 
$f(\tilde{\epsilon}) = 1/[1 + e^{(\tilde{\epsilon}-\mu)/k_BT}]$, proceeding along this line 
until self-consistent convergence.

It is easy to see that the equilibrium occupancy $N-\mu$ \cite{rbhasko2} with respect to chemical potential (we drop the angular term indicating
average here) should be qualitatively different between the SCF and 
many-body results. In the former, the electron occupancy is a fractional amount, adiabatically changing from zero 
to two. In the many-body result, however, one does not simply multiply the results for one electron by two, but the 
electron occupancy
changes abruptly between zero to one, followed by a plateau of width $U$ over which the electrons
are blockaded by the Coulomb interaction, after which the electron number reaches two
abruptly. 

One could capture this Blockaded effect using an {\it{unrestricted}} self-consistent potential (USCF) by 
dictating that the up and down spins do not feel potentials due to themselves. By eliminating this self-interacting,
the unrestricted potential for a particular spin is then given by 
\begin{eqnarray}
U_{USCF} &=& \langle \partial H_{int}/\partial \hat{n}_\sigma\rangle \nonumber\\
&=& U n_{\bar{\sigma}}\nonumber\\
&=& U(N-n_\sigma)
\label{uscf}
\end{eqnarray}
with $n = \langle \hat{n}\rangle$ and $\bar{\sigma}$ represents the spin opposite to
$\sigma$. This spin-dependent unrestricted potential eliminates the self-interaction of the level to which charge 
is being added. A self-consistent solution of the occupancy yields an $N-\mu$ \cite{rbhasko2} plot very similar to the exact result, 
showing that an unrestricted calculation can capture {\it{equilibrium}} Coulomb Blockade effects. 

\subsection{B. Single level system, Fock space transport}
Non-equilibrium turns out to be hard to mimic with any SCF theory, even with considerable 
latitude in our choice of the SCF potential. Let us assume the contact injection rates are given by
$\gamma_{1,2}/\hbar$, ignoring level broadening for the moment. Here 1,2 also refers to the left (L) and right (R) contact. Both conventions are used in this paper to be consistent with NEGF literature. One can write down a master equation for a 
transition between the many-electron levels driven by the contacts
\begin{equation}
\dot{P_i} = -\sum_j R_{ij}P_i + \sum_j R_{ji}P_j
\end{equation}
where $i,j$ represent the many-body states. The master equation,
intuitively quite transparent from Figure~\ref{fig_2}(b), can be formally derived by decoupling the contact 
and molecular density matrix equations in the steady-state limit and then invoking a Markov approximation 
that ignores memory effects such as energy-dependences in the contact broadening. For our simple example
of the dot with two spin levels, the transition rates between the four Fock
states, numbered as $\{|00\rangle, |01\rangle, |10\rangle, |11\rangle\}$, are given by
\begin{eqnarray}
R_{00\rightarrow 01} &=& R_{00\rightarrow 10} = (\gamma_1f_1 + \gamma_2f_2)/\hbar\nonumber\\
R_{01\rightarrow 00} &=& R_{10\rightarrow 00} = (\gamma_1\bar{f}_1 + \gamma_2\bar{f}_2)/\hbar\nonumber\\
R_{01\rightarrow 11} &=& R_{10\rightarrow 11} = (\gamma_1{f^\prime}_1 + \gamma_2{f^\prime}_2)/\hbar\nonumber\\
R_{11\rightarrow 01} &=& R_{11\rightarrow 10} = (\gamma_1\bar{f^\prime}_1 + \gamma_2\bar{f^\prime}_2)/\hbar
\end{eqnarray}
where $f_{1,2} = 1/[1 + e^{(\epsilon_0-\mu_{1,2})/k_BT}]$, $\bar{f}_{1,2} = 1-f_{1,2}$, $f^\prime_{1,2} 
= 1/[1 + e^{(\epsilon_0+U-\mu_{1,2})/k_BT}]$, and $\bar{f^\prime}_{1,2} = 1-f^\prime_{1,2}$. In short, electron
addition processes are governed by probability of occupancy $f$ at the corresponding transition energies $\epsilon$
or $\epsilon + U$ by each
contact electrochemical potential, while electron removal processes are governed by the probability of vacancy $1-f$. 
Thus the two current onset points occur at the bias situations shown schematically in Figure~\ref{fig_2}(c). 

At steady-state, it is straightforward to solve these equations (only three of which are independent), along with the
normalization $\sum_iP_i = 1$. The probabilities are then used to calculate the
current injected by one contact (say the left one, ``L') as
\begin{equation}
I^L = \sum_i (\pm e/\hbar)[-\sum_j R^L_{ij}P_i + \sum_j R^L_{ji}P_j]
\end{equation}
where the rates $R^L$ are obtained by only considering the individual left contact 
contributions to the corresponding rate, for example, 
$R^L_{00\rightarrow 10} = \gamma_1f_1/\hbar$. The $\pm$ signs correspond to
addition/removal of electrons by the left contact. The resulting I-V characteristic is 
shown in Fig.~\ref{fig_2}(d).

In an SCF treatment, the current, shown dotted in Figure~\ref{fig_2}(d) is obtained by 
solving the rate equations in the one-electron subspace. The electron occupancy is
given by $N = (\gamma_1f_1 + \gamma_2f_2)/(\gamma_1 + \gamma_2)$, where the Fermi functions 
are evaluated at the energy $\tilde{\epsilon} = \epsilon + U_{SCF}$. The SCF potential in 
turn depends on $N$ as described earlier, so the calculation is done self-consistently. The 
converged energy is then used to calculate the current as $I = (2e/\hbar)
\gamma_1\gamma_2/(\gamma_1+\gamma_2)[f_1-f_2]$. 

The RSCF model that treats spins equally tends to give an adiabatically increasing current 
that reaches its maximum contact-dominated value $2e/\hbar\times \gamma_1\gamma_2/
(\gamma_1+\gamma_2)$ when the contact electrochemical potential fully crosses the level. It 
is important to note that charging alone can smear out this current, leading to a 
low conductance value spread out over a wide voltage range comparable to $U$. The RSCF 
potential $U(N-1/2)$ causes a continuous shift in levels with charge addition, which is 
accomplished in fractions. The unrestricted approach USCF gives an intervening Coulomb
Blockade plateau of width $U$ that separates the first spin addition (or removal) 
event from the second. The intervening `open shell' plateau is at half the maximum value 
for complete level filling, which is understandable because the two spins are treated on
an equal footing chemically, and therefore carry equal current. 

Compared with the SCF results above, the exact solution
of the many-body rate equations reveals an interesting surprise that is actually quite 
illuminating. Of all many-electron configurations, the 1-electron states (and only those)
are doubly degenerate, giving us a normalization condition that differs from a simpler
version that ignores spins and simply multiplies all results by two. As a consequence of
this sum-rule, which takes Pauli exclusion into account (preventing double up or down 
spin states, for example), the exact value of the open-shell current plateau depends on 
the Fermi functions as $\gamma_1\gamma_2/[\gamma_1(1+f_1) + \gamma_2(1+f_2)]$ in the large
charging ($U\rightarrow \infty$) limit. For sharp levels at positive bias ($f_1 = 1$, $f_2
= 0$), this reaches $\gamma_1\gamma_2/[2\gamma_1 + \gamma_2]$. For the strongly non-
equilibrium situation corresponding to equal resistive couplings ($\gamma_1 = \gamma_2$), the 
Coulomb plateau carries two-third of the maximum closed-shell current, in contrast with the 
USCF result that gives a factor of half. This implies an interesting history dependence, in 
that the first spin added to the empty level carries more current than the second! If we keep 
track of the entire many-electron configuration space, it is easy to see that this counter-
intuitive result arises because there are two ways of adding the first spin and only one way 
of adding the second (the other channel eliminated by exclusion). This subtlety is completely 
washed away when we choose to work in a reduced $N\times N$ (or $2N\times 2N$ for unrestricted) subspace 
instead of the full $2^{2N}\times 2^{2N}$ configuration space. 

The SCF potential $\langle\partial H/\partial n\rangle$ was calculated by writing the electron operators $\hat{n} = 
\langle \hat{n}\rangle + \delta\hat{n}$, expanding the Coulomb term $U\hat{n}_\uparrow\hat{n}_\downarrow$ and dropping the 
correlation terms $\delta\hat{n}_\uparrow\delta\hat {n}_\downarrow$ completely, i.e., the Hartree-Fock approximation.
One could include parts of the correlation term
phenomenologically, by dictating that $\hat{n}_i\hat{n}_j \approx (1-g_{ij})\langle n_i\rangle\langle n_j\rangle$, with 
$g_{ij}$ representing the exchange-correlation hole. This is in the spirit of Kohn-Sham theory, where the 
potential is calculated by various approximate means. However, the effect of $g$ is simply to renormalize the 
charging energy $U$ that
it adjoins, influencing at best the {\it{width}}, but not the {\it{height}} of the 
current plateaus. Thus, {\it{even for the simplest quantum dot, unrestricted potentials in 
the one-electron subspace cannot capture the nonequilibrium properties correctly}}.

\begin{figure}[ht]
\centerline{\epsfig{figure=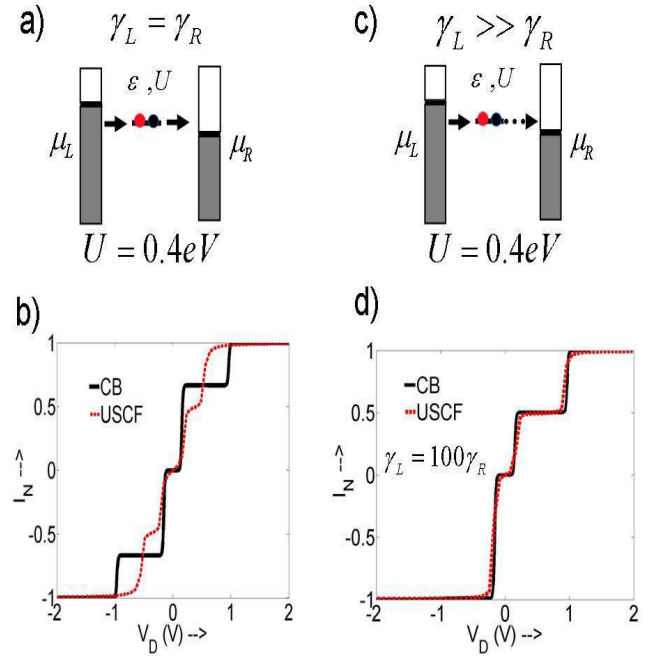,height=3.5in,width=3.5in}}
\caption{Comparison between USCF and exact transport results for (a) extreme non-equilibrium
($\gamma_1 = \gamma_2$), b) in which the I-V's show significant discrepancy between USCF (dotted red) and exact (bold black) result. c) The case of strong asymmetry implies a  near equilibrium with the first contact ($\gamma_1 = 100
\gamma_2$). d) The discrepancies in plateau onset, width and height are resolved in this
equilibrium limit, but are hard to resolve once we go far from equilibrium.} 
\label{fig_3n}
\end{figure}

It is further illuminating when a comparison between the USCF and exact model is performed. Recall that in the USCF approach, we introduce the aspect of ``self interaction correction" as described 
in Eq.~\ref{uscf}. Here, the presence of an electron of a particular spin adds a coulomb cost to the other, but not on itself. 
Fig.~\ref{fig_3n}(b) shows the discrepancy between the USCF and the many-body transport calculations for a dot with two spin levels, coupled equally to two contacts (driving it 
far from equilibrium). As is clear from the results, the discrepancy is in the onset voltages, widths and heights
of the various plateaus. While the plateau widths could be adjusted to fit the exact results
by renormalizing the charging energies parametrically to account for correlation effects, 
the heights are independent of these values, and depend only on universal factors arising
from Pauli exclusion, and cannot be fixed in such a straightforward way, or by the usual
DFT approach of progressively improving on correlations in the electronic structure. The
issue is further underscored by the fact that in the strongly asymmetric limit ($\gamma_1 
= 100 \gamma_2$), where the system is essentially driven into equilibrium with the left
contact, the agreement between USCF and many-body results is substantially improved. It
is worth clarifying at this time though that in the multilevel generalization, this 
correspondence becomes a lot harder to establish {\it{even near equilibrium}}, since even
the number of current plateaus generated by a $2^N \times 2^N$ Fock space approach differs 
substantially from its $2N \times 2N$ USCF counterpart (unless the broadening functions
bear enough poles through their energy-dependences to precisely account for those missing
conductance peaks). 

The situation gets simpler if we have asymmetric contacts $\gamma_1 \gg \gamma_2$, so that
the system approaches equilibrium with the left contact. As Fig.~\ref{fig_3n} (d) shows, the USCF agrees
with the many-body limit in this asymmetric coupling case, as both approaches are dealing with
a near-equilibrium problem. For positive bias on the weaker contact, 
the stronger contact keeps the level filled (which it can do in two ways, adding an up or a 
down spin, assuming the level was empty to begin with). For opposite bias, the stronger 
contact empties this level, which can now be done in only one way (up OR down depending on
what occupied the level). The competition between `shell tunneling' and `shell filling' 
makes the I-V strongly asymmetric, with the first plateau half the second
for positive bias on the weaker contact, and merging with the second for opposite bias. 
This ratio of one to two, observed experimentally \cite{rralph2}, arises in a straightforward 
way from our analyses since the ratio of the first and second plateau currents is given for 
positive bias by $(\gamma_1+\gamma_2)/(2\gamma_1+\gamma_2) \approx 1/2$ for $\gamma_1
\gg\gamma_2$, and by $(\gamma_1+\gamma_2)/(\gamma_1+2\gamma_2) \approx 1$ for negative bias. 
The asymmetry arises from the difference in the number of spin addition and removal channels 
for positive and negative bias, and leads to an asymmetry in the current levels.

We will next show how this model is extended for a larger molecule, and how additional 
physics due to correlations and excitations start to arise. 

\subsection{C. General Approach for multilevel systems}
In case of a larger molecule, one begins with the model Hamiltonian in second quantized notation:
\begin{eqnarray}
\hat{H} &=& \sum_{\alpha} \epsilon_{\alpha} \hat{n}_{\alpha}
+ \sum_{\alpha \neq \beta} t_{\alpha \beta} c_{\alpha}^{\dagger} {c_\beta} \nonumber\\
&+& \sum_{\alpha,\sigma} U_{\alpha \alpha} \hat{n}_{\alpha\sigma}
\hat{n}_{\alpha\bar{\sigma}} + \frac{1}{2} \sum_{\alpha \neq \beta} U_{\alpha \beta}
\hat{n}_{\alpha} \hat{n}_{\beta} ,
\label{eq:embh}
\end{eqnarray}
where $\hat{n}_\alpha = c^\dagger_\alpha c^{}_\alpha$, $\alpha,\beta$
correspond to the orbital indices of the orbitals for various sites on the molecule, 
and $\sigma$,$\bar{\sigma}$ represent a particular spin and its reverse. Exact 
diagonalizing this Hamiltonian yields a large spectrum of closely spaced excitations 
in every charged molecular configuration. Using the equation of motion of the 
density matrix of the composite molecule and leads and assuming no molecule-lead
correlations, one can derive \cite{timm,rbraig} a simple master
equation for the density-matrix of the system. Ignoring off-diagonal
coherences, we are left with a master equation \cite{rbraig} in terms
of the occupation probabilities $P^N_i$ of each N electron many-body
state $|N,i\rangle$ with total energy $E^N_i$. The master equation then
involves transition rates $R_{(N,i)\rightarrow(N\pm 1,j)}$ between
states differing by a single electron, leading to a set of independent
equations defined by the size of the Fock space \cite{rralph}
\begin{equation} 
\frac{dP^N_i}{dt} =
-\sum_{N,j}\left[R_{(N,i)\rightarrow(N\pm 1,j)}P^N_i - R_{(N\pm
1,j)\rightarrow(N,i)}P^{N\pm 1}_j\right] 
\label{ebeenakker}
\end{equation} 
along with the normalization equation $\sum_{i,N}P^N_i =
1$. For weakly coupled dispersionless contacts, parameterized using
bare-electron tunneling rates $\gamma_{\alpha}$, ($\alpha$: left/right
contact), we define rate constants 
\begin{eqnarray}
\Gamma_{ij\alpha}^{Nr} &=& \gamma_\alpha|\langle
N,i|c^\dagger_\alpha|N-1,j\rangle|^2\nonumber\\ \Gamma_{ij\alpha}^{Na}
&=& \gamma_\alpha|\langle N,i|c^{}_\alpha|N+1,j\rangle|^2,
\end{eqnarray} 
$c^\dagger_\alpha,c^{}_\alpha$ are the
creation/annihilation operators for an electron on the molecular end
atom coupled with the corresponding electrode. The transition rates
are given by 
\begin{eqnarray} 
R_{(N,i)\rightarrow(N-1,j)} &=&
\sum_{\alpha=L,R}\Gamma_{ij\alpha}^{Nr}\left[1-f(\epsilon^{Nr}_{ij}-\mu_\alpha)\right]
\nonumber\\ R_{(N-1,j)\rightarrow(N,i)} &=&
\sum_{\alpha=L,R}\Gamma_{ij\alpha}^{Nr}f(\epsilon^{Nr}_{ij}-\mu_\alpha).
\end{eqnarray} 
for the removal levels $(N,i \rightarrow N-1,j)$, and
replacing $(r \rightarrow a, f \rightarrow 1-f)$ for the addition
levels $(N,i \rightarrow N+1,j)$. $\mu_\alpha$ are the contact
electrochemical potentials, $f$ is the corresponding Fermi function,
with single particle removal and addition transport channels
$\epsilon^{Nr}_{ij} = E^N_i - E^{N -1}_j$, and $\epsilon^{Na}_{ij} =
E^{N+1}_j - E^N_i$. Finally, the steady-state solution to
Eq.(\ref{ebeenakker}) is used to get the left terminal current
\begin{equation} 
I =
\pm\frac{e}{\hbar}\sum_{N,i,j}\left[R^L_{(N,i)\rightarrow(N\pm 1,j)} P^N_i
- R^L_{(N\pm 1, j)\rightarrow(N,i)}P^{N\pm 1}_j \right] 
\end{equation}
where states corresponding to a removal of electrons by the left
electrode involve a negative sign. We will assume a break-junction configuration
with equal electrostatic coupling with the leads, $\mu_{L,R} =
E_F \mp eV_d/2$. 

While the above equations include spectral details from the multiple excitations,
a considerable simplification arises if we can incoherently sum over many of these
excitations, leading to the `orthodox model', where
\begin{equation} 
I =
\pm\frac{e}{\hbar}\sum_{N}\left[R^L_{N\rightarrow N\pm 1} 
- R^L_{N\rightarrow N\mp 1}\right]P^{N} 
\end{equation} 
The transition energies can be obtained in terms of a simple RC circuit, while
the transition rates also compactify once we integrate them over the relevant
energies, taking exclusion factors into account.  

The implementation of the above sets of equations now sets the stage for further 
discussions.

\section{III. Coulomb Blockade: the case of electronic transport channels}
In this section we describe how the excitation spectra of molecules may be probed 
using Coulomb Blockade transport. While typical molecular I-Vs look relatively 
featureless, what is not often appreciated is that in the Coulomb Blockade limit 
(realized by engineering weak contacts or non-conductive backbones), electronic 
excitations do give rise to prominent features in the I-V \cite{jpark,zhitenev}. 
We will also show how such an excitation spectra can generate {\it{intrinsic}} 
I-V asymmetries, which in turn can manifest itself as NDR effects. 

\begin{figure}
\centerline{\epsfig{figure=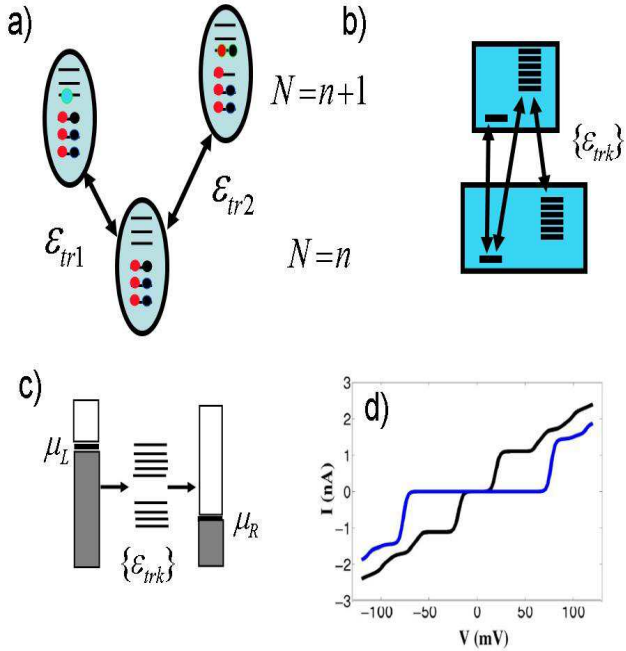,height=3.5in,width=3.5in}}
\caption{Electronic excitations in molecular conduction. a) Schematic shows three Fock states, the $N=n$ equilibrium ground state configuration and two states
in the addition $N=n+1$ spectrum. b) Each charge configuration consists of a ground state and a set of closely spaced excitations. The access of such closely spaced excitations 
between the neutral and charged species gives rise to numerous c) transport channels $\epsilon_{trk}$ \cite{rbhasko1}. The access of such transport channels 
gives rise to a quasi ohmic behavior in the I-V characteristics that are commonly noted in various experiments \cite{rweber1,rweber2}. } 
\label{fig_3}
\end{figure}

Exact-diagonalization of the molecular Hamiltonian Eq.~\ref{eq:embh} generates a 
body of excitation spectra for every charged configuration. Our typical starting point
is the equilibrium molecular quantum dot configuration in its ground state. Addition or 
removal of an electron takes the system to the a new ground state corresponding to
the singly charged cation or anion, marking the onset of conduction. In addition to the 
ground state however, each charged species bears a quasi continuous excitation spectrum 
separated from its ground-state energy by a gap that is determined by the energetics of 
the molecular system. Once the 
first excitation is accessed, the quasi-continous excitation spectrum can be easily 
probed, as shown in Figures~\ref{fig_3}(a), (b) and (c). It is worth emphasizing that
the excitations are energetically quite close as they differ in the rearrangement (but 
not in the number) of charges. At high bias with small broadening and in the absence of
phonons, the purely electronic excitations do not have adequate time to relax to the ground 
state, accounting for their visibility in the current spectrum. 

The simplest and most prominent impact of Coulomb Blockade on the I-Vs of 
short molecular wires is a clear suppression of zero-bias conductance, often seen 
experimentally \cite{rreed,rtao}. 
However, integer charge transfer can also occur between various electronic
{\it{excitations}} of the neutral and singly charged species at
marginal correlation costs \cite{rfulde,gus}. The above fact leads to fine
structure in the plateau regions \cite{jpark,rweber1,rweber2,pnas},
specifically, a quasilinear regime resulting from very closely spaced
transport channels ($\epsilon^{N}_{ij}$) via excitations. The crucial
step is the access of the first excited state via channel
$\epsilon^{Nr}_{10}$, following which transport channels involving
higher excitations are accessible in a very small bias window. 

The sequence of access of transport channels upon bias, enumerated in 
the state transition diagrams shown in Figs.~\ref{fig_3}(a),(b) and 
(c) determines the shape of the I-V characteristic. 
When the Fermi energy $E_{F}$ lies closer to the
threshold transport channel $\epsilon^{Nr}_{00}$ corresponding to charge
transfer between two ground states, it takes an additional positive drain 
bias for the source to access the first excited state of  the neutral 
system via the transition $\epsilon^{Nr}_{10}$. Under this condition,
the I-V shows {\it{a sharp rise followed by a plateau}} (Figure~\ref{fig_3}(d), 
shown in blue), as seen in various experiments \cite{rdekker}. However, 
when transport channels that involve low lying excitations such as 
$\epsilon^{Nr}_{10}$ are closer to the Fermi energy $E_F$ than 
$\epsilon^{Nr}_{00}$,
the excitations get populated by the left contact immediately when the
right contact intersects the threshold channel $\epsilon^{Nr}_{00}$,
allowing for a {\it{simultaneous}} population of both the ground and
first excited states via $\epsilon^{Nr}_{00}$ and $\epsilon^{Nr}_{10}$
at threshold. Under these conditions the I-V shows {\it{a sharp onset
followed immediately by a quasilinear regime}} (Fig.~\ref{fig_3}(d) 
shown in black) {\it{with no
intervening plateaus}}, as observed frequently in I-Vs of molecules
weakly coupled with a backbone \cite{rweber1,rweber2,jpark}.
\begin{figure}
\centerline{\epsfig{figure=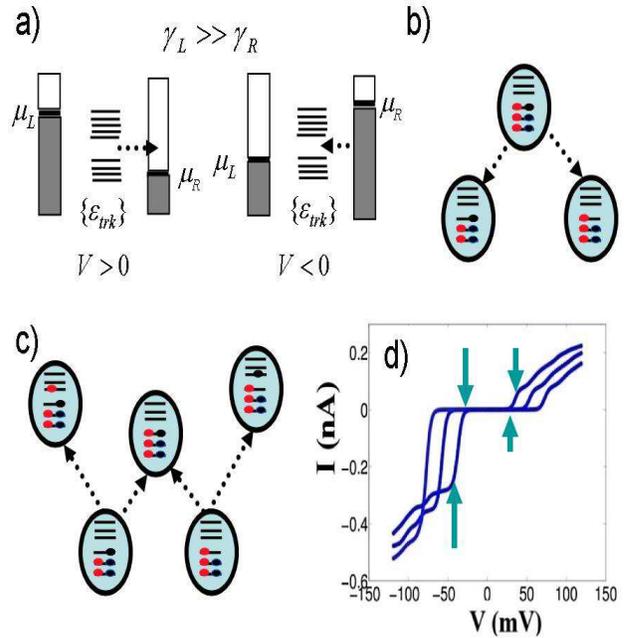,height=3.5in,width=3.5in}}
\caption{Current asymmetry from extrinsic ``contact" asymmetry. a) In the case when the left electrode is more strongly coupled to the channel than the right one $\gamma_L >> \gamma_R$, currents are limited by removal (addition) rates under forward (reverse) bias conditions. b) The number of removal channels can be significantly different from the number of c) addition channels. This gives rise to the asymmetry between d) forward and reverse bias situations. Furthermore the application of gate potential modulates the number of transport channels available at threshold. This feature results from the Fock space excitation spectra, causing not only a change in threshold voltage with applied gate bias, but also a discernable threshold current modulation as noted here, and in various experiments \cite{jpark}.} 
\label{fig_4}
\end{figure}
\subsection{A. Extrinsic Asymmmetries -the case of rectification}
The direct role of excitations in conduction becomes particularly
striking under asymmetric coupling ($\gamma_L>>\gamma_R$) with
contacts \cite{jpark,rscott}. Due to this extrinsic asymmetry, 
current magnitudes are dictated by the weaker contact. This asymmetry 
directly affects the forward and reverse bias characterestics, leading 
to current rectification. This rectification is caused by the inherent 
asymmetry between addition and removal processes, each of which is the 
rate limiting depending on the bias direction, as shown in 
Figs.~\ref{fig_4}(a), (b) and (c).

In contrast to the SCF regime where
unequal charging drags out a same level current over different
voltage widths \cite{rasymm}, in the CB regime we encounter clear
intermediate current steps from open shells, with current heights 
that are themselves are asymmetric at threshold (Fig.~\ref{fig_4}(d)). 
This asymmetry arises due to the difference in the number of pathways for
removing or adding a spin, taking in particular into account the possible
excitation channels between the neutral and singly charged species
(Figures~\ref{fig_4}(b) and (c)). The number of such accessible excitations at threshold
can be altered with an external gate bias, leading to a prominent gate
modulation of the threshold current heights, over and above the
modulation of the onset voltages and the conductance gap \cite{jpark}
(Figure~\ref{fig_4}(d)). Furthermore, it is easy to show that the asymmetry will 
flip between gate voltages on either side of the charge degeneracy point,
as is also observed experimentally \cite{rscott}. 

It is worth emphasizing that the sophistication arose specifically due to the 
presence of separate identifiable contributions from the open shells, which 
are normally overwhelmed by broadening if we were away from the CB limit. However,
the specific identification of these excitations is not crucial to the qualitative
shape of the I-Vs, as long as one is not looking too closely at the individual
spectral features. A simpler orthodox model would then suffice to reproduce these
broad features described above, such as the transition between steps and slopes
in the I-Vs, the flipping of asymmetry and the gate modulation of the current 
levels \cite{rowen}. However, there are notable exceptions, such as intrinsic
asymmetries, where a clear separation needs to be made between certain classes of
excitations with longer lifetimes (`traps') and the regular excitations with 
shorter lifetimes (`channels') that are responsible for the current flow. 

\subsection{B. Intrinsic Asymmetries - NDR, hysteresis and telegraph noise}

The physics of NDR can be explained readily using an USCF model that is 
actually quite intuitive. Consider a channel and a trap, the channel being
strongly coupled to the contacts and the trap weakly coupled. Accessing the channel
creates a resonant onset of current. Accessing the trap subsequently would keep
the conduction unaffected, as the trap is non-conducting; however, once we throw
in the strong Coulomb repulsion arising from charging up the trap, it is easy to 
see that this repulsion can expel the conducting channel out of the conducting 
bias window, leading to an NDR. Self-interaction correction is crucial to this 
picture, as the charging should repel the channel level, but not the trap level
itself. Further sophistications can arise by considering the lifetime of the trap.
If the trap does not release its charge during the measurement time, then a
reverse scan would keep the channel blocked and lead to a hysteresis. Such a
hysteresis is scan-rate dependent, as the rate determines the degree to which
the trap releases its captive charge. The low current state can be reversed
by going to large negative bias to expel this charge and restore initial 
conditions. 

The realization of the NDR involves two conditions: (a) that the charging of the
trap is large enough to expel much of the channel from the bias window, and 
(b) that the resulting current carried by the trap plus the residual tail of the
channel sitting in the bias window exceeds the current carried originally by
the channel alone. The first amounts to the onset condition for an NDR, while
the second describes, in some sense, the effectiveness of the NDR (in terms of
an inequality). With a little bit of effort, one can extract a range of parameters
that satisfy both these conditions. We will instead show that even in the Fock
space picture (which treats the channel plus trap as a composite system), one
can identify these two same conditions: the NDR starts when one encounters what
we refer to as a `dark' or `blocked' state, and subject to this condition, the
effectiveness of the NDR amounts to an inequality that we will now discuss. 

\begin{figure}
\centerline{\epsfig{figure=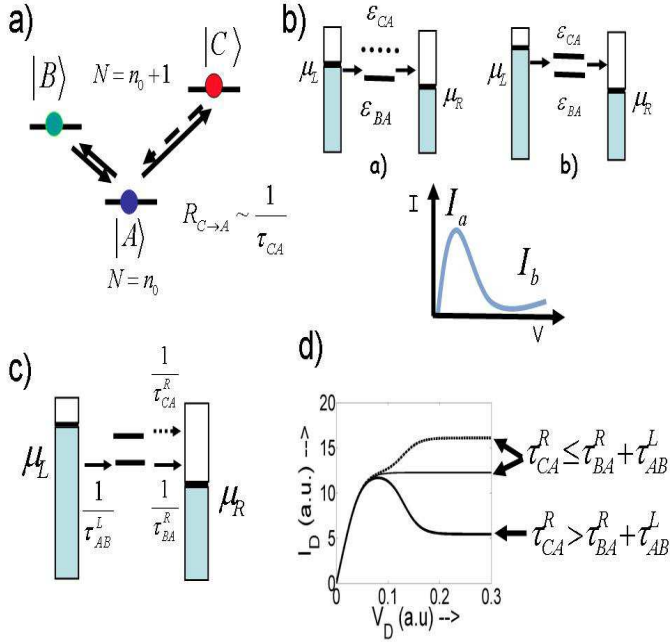,height=3.5in,width=3.5in}}
\caption{NDR effects from intrinsic asymmetry or ``dark states". a) A generic mechanism 
NDR can be cast in terms of three device Fock-space states, and transitions $\mid A 
\rangle \leftrightarrow \mid B \rangle$ and $\mid A \rangle \leftrightarrow \mid C 
\rangle$ between those that differ by a single electron. The dark state $\mid C \rangle$ 
say has a very slow removal rate in comparison to its addition rate. b) Bias environments, 
a and b, correspond to current rise followed by a collapse. c) A general criteria for 
such a current collapse or NDR to occur with increasing voltage can be cast in terms 
of the dark state removal rates. d) NDR in the I-V characteristic is contingent to the 
criteria derived here \cite{rbhasko3}.} 
\label{fig_5}
\end{figure}

A condition for NDR can be derived much more generally in terms of three Fock space 
states $|A \rangle$, $|B \rangle$, and $|C \rangle$ with energies $E_A<E_B<E_C$ 
respectively (Fig~\ref{fig_5}(a)), representing three accessible states 
within the bias range of interest. For instance, $|A \rangle$, $|B \rangle$ could 
be the ground states of the $N=n_0$ and $N=n_0+1$ electron systems, while $|C 
\rangle$, the first excited state of the $N=n_0+1$ electron system. Transport of 
electrons involves single charge removal or addition between states $|B\rangle$, 
$|C \rangle$ and $|A \rangle$ that differ by an electron, via addition and 
removal rates $R_{A \leftrightarrow B,C} \propto \frac{1}{\tau_{AB,C}}$. Such an 
electron exchange is initiated when reservoir levels are in resonance with the single 
electron transport channels $\epsilon_{BA}=E_B-E_A$ and $\epsilon_{CA}=E_C-E_A$ 
respectively. The I-V characteristic of this three state system, shown schematically 
in Fig~\ref{fig_5}(b), shows two plateaus with current magnitudes $I_{a}$ and $I_{b}$ 
respectively. Current collapse or NDR occurs when $I_{a} > I_{b}$. 

In the above system, NDR occurs when under specific conditions, the state $|C \rangle$ can 
be a blocking or dark state, for which electron addition is feasible while removal 
is rate limiting. This can happen when there are intrinsic asymmetries within the transport 
problem, as in the trap-channel dichotomy described earlier. Regardless of the specific
origin of blocking, a simple criteria for current collapse can be derived \cite{rbhasko3} 
in terms of the rate limiting process, that is the electron removal rate 
$R_{C \leftrightarrow A}$ from the dark state $|C \rangle$. A better intuition is provided 
by thinking of rates as inverse lifetimes, as shown in Figure~\ref{fig_5}(c). 
If the life time of state $|C \rangle$, $\tau_{CA}$ exceeds the sum of the addition and 
removal rates of the conducting state, it becomes an effective blocker. In other words, the 
condition $\tau^R_{CA} > \tau^L_{AB} + \tau^R_{BA}$ determines whether NDR occurs or not. 
The superscripts $L$ and $R$ represent left or right electrode, and under the forward bias 
situation add and remove electrons respectively. The consequence of the above criterion is 
summarized in Figure~\ref{fig_5}(d), clearly indicating that NDR only occurs when 
$\tau^R_{CA} > \tau^L_{AB} + \tau^R_{BA}$. A similar criteria is valid for the reverse 
bias direction by interchanging superscripts $L$ with $R$. 
\begin{figure}
\centerline{\epsfig{figure=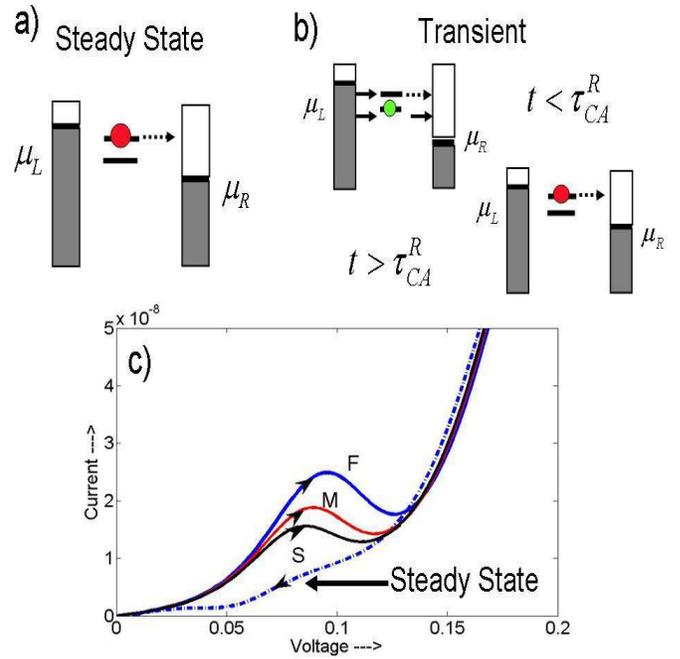,height=3.5in,width=3.5in}}
\caption{Hysteresis effects due to sweep rate. a) Under steady state conditions the dark state behaves as an effective blocker. b) A sweep rate faster 
than the rate determining time scale can put the transport process in a transient state. c) Under such transient conditions sweep rate can induce disguised NDRs and often 
hysteresis, both depending on sweep rate \cite{kiehl}. } 
\label{fig_6}
\end{figure}

Further sophistications may arise due to `transient' probing. The criteria obtained above applies for steady state, 
where the dark state gets occupied with certainty, resulting in an inevitable current blockade regimee. 
Let us consider such a blocked current state, shown dotted in Figure~\ref{fig_6}(c). A `disguised' NDR can also be 
achieved if the voltage sweep rates are faster than the rate determining dark state life time. Such a sweep rate 
simply means that the dark state has not yet been occupied and thus a current determined by the addition and removal 
times ${1}/({\tau^L_{AB} + \tau^R_{BA}})$, can still flow. This situation is shown in Figure~\ref{fig_6}(c), in 
which the onset of dark state can be delayed provided the sweep rate is fast enough. But once the 
dark state forms, the current is blocked and thus remains so. This implies that on the backward sweep current 
remains blocked, resulting in a sweep rate-dependent hysteresis. Such a behavior has been noted by Kiehl 
{\it{et.al.}} \cite{kiehl} in a molecular system, which was attributed to a slow charge trapping process, which again 
fits into the dark state picture. Just as noted here, Kiehl {\it{et.al.}} also observe that the NDR effect diminishes 
with decreasing sweep rate and ultimately vanishes at steady state, implying the absence of a real hysteresis at 
steady state. This kind of hysteresis is just an artifact of sweep rates being faster than the rate determining 
processes. 

It must be mentioned that certain hysteretic processes can occur even at steady state due to the true bistable 
nature of the system. A classic example is the charging induced hysteresis in resonant tunneling diode (RTD) 
structures \cite{cap}. Other interesting examples include bistability caused by hyperfine interactions \cite{tar2} 
causing a hysteresis with applied magnetic field, and optical bistability created by superlattice dielectrics
that show strong terahertz nonlinearity due to Bloch oscillation and dynamic localization of electrons \cite{thz1,
thz2}. 
        
While we discussed the interesting case of dark states earlier, touching upon relevant consequences such as NDR 
and hysteresis, we avoided particular examples. Recently, we applied our dark state model \cite{rbhasko3} to 
explain the NDR observed due to subtle spin correlation effects in double quantum dots \cite{tar}. Indeed this 
NDR results in the spin blockade regime \cite{tar} which currently forms a key concept in the area of single spin 
manipulation and control \cite{kow,pet,cm,kw,now}. 

Finally, the transient probing with a high resolution probe would allow us to explore the approach to resonance
with the trap states. As we discuss in \cite{ghoshieee}, the stochastic blocking and unblocking of the channel
by the occupation/deoccupation of the traps near resonance generates a flicker in the output current known as
random telegraph noise. The ratio of capture and emission times is given by a Boltzmann factor whose argument
depends on the trap energy location. Shifting this trap with a combination of gate and drain voltages leads to
an associated scaling of the capture to emission time ratio, allowing one to infer the spatial and spectral 
location of the trap states. This provides a powerful `barcode' for characterizing single-molecular defects. 

Our analyses over the last few sections focused on electron-electron correlations and their treatment in a Fock
space approach that captures the relevant physics (if at the expense of computational simplicity). In the next
section, we will discuss how this approach can be extended to incorporate electron-phonon correlations in transport.

\section{IV. Electron-phonon Interactions:}
So far, our primary concern was Coulomb Blockade and electronic excitations. Here, we show that coupling Coulomb 
Blockade with phonon-assisted tunneling results in non-trivial physics. This section is primarily motivated by a 
recent series of experiments performed on suspended carbon nanotube quantum dots \cite{leroy}. In such a suspended 
system, the phonons are driven far out of equilibrium and couple strongly with the electronic system. Our approach 
then is to invoke the Fock space of the electron phonon system as indicated in Figures~\ref{fig_1}(c) and (d). 

\subsection{A. SCF treatment -- IETS and phonon sidebands}
The SCF approach to electron-electron interactions has been discussed at length in our earlier papers and
contrasted with the Fock space approach. It is worth quickly touching upon the SCF treatment of phonons, 
before diving into its Fock-space analogue. In presence of dephasing scattering events, the current at the left
contact can be written in the NEGF formalism \cite{rdatta1} as 
\begin{equation}
I_L = {{2e}\over{h}}\int dE Tr[\Sigma_L^{in}A-\Gamma_LG^n]
\label{inegf}
\end{equation}
where $A$ is the spectral function, $\Gamma_L$ is the broadening by the left
contact, $\Sigma_L^{in}$ is the in-scattering self-energy from the left contact
and $G^n$ is the correlation function describing the energy-dependent occupancy
of the levels, taking quantum interference into account. The influence of scattering
by contacts and phonons sits in $\Sigma^{in} = \Sigma^{in}_L + \Sigma^{in}_R + 
\Sigma^{in}_{ph}$. The equations for the contact $\Sigma$s are well known  -- we will
focus here on the additional contributions from the phonon scattering processes. 
Within the self-consistent Born approximation (self-consistency 
needed to conserve current), the equations connecting the phonon contributions form 
two groups -- a set of dynamic equations, and a set of static equations. The non-equilibrium 
dynamics describing the filling and emptying of states is described by
\begin{eqnarray}
G^{n,p}(E) &=& G(E)\Sigma^{in,out}_{ph}(E)G^\dagger(E) \nonumber\\
\Sigma^{in,out}_{ph}(E) &=& D_0(\omega)\otimes\Biggl[n_B(\omega)G^{n,p}(E\mp\hbar\omega) \Biggr.\nonumber\\
\Biggl. &+& \left(n_B(\omega)+1\right)G^{n,p}(E\pm\hbar\omega)\Biggr]
\label{rkeldys}
\end{eqnarray} 
where $n_B(\omega) = [e^{\hbar\omega/k_BT}-1]^{-1}$ is the equilibrium Bose-Einstein distribution of the phonons,
and $D_0$ is its deformation potential, in other words, $(D_0)_{ij} = h_ih_j$, $h_{\alpha}$ being the electron phonon
coupling constant at the real space point $\alpha$ (In general, $h$ is a non-diagonal matrix in an arbitrary basis set,
and $D_0$ is a fourth-rank tensor). $\otimes$ denotes an element-by-element (as opposed to matrix)
multiplication. The equations above are for a single phonon mode at frequency $\omega$, and will
need to be integrated over a phonon density of states for a continuous distribution of phonons. 

The static equations describing the states themselves are given by
\begin{eqnarray}
\Gamma_{ph}(E) &=& \Sigma^{in}_{ph}(E) + \Sigma^{out}_{ph}(E)\nonumber\\
\Sigma_{ph}(E) &=& {\cal{H}}(\Sigma_{ph})+i\frac{\Gamma_{ph}}{2}\nonumber\\
G(E) &=& [EI - H - \Sigma_L - \Sigma_R - \Sigma_{ph}]^{-1},
\end{eqnarray}
where ${\cal{H}}$ denotes the Hilbert transform.
The above equations capture the physics of phonon-assisted tunneling in all its various limits. 
In its simplest form, it contributes to dephasing that reduces the ballistic current in devices. 
Near resonance, the $\pm \hbar\omega$ terms in the arguments of the $G^{n,p}$ matrices contributing
to $\Sigma^{in,out}(E)$ generate phonon sidebands. The scaling of these sidebands depends on the
deformation potential as well as the distribution of the phonons. We have assumed this to be equilibrium
Bose-Einstein $n_B$, although one could generalize it provided we have a separate evolution equation
for the phonon dynamics coupled to the electron transport equations that describe parametrically or
otherwise how the phonons are driven away from equilibrium by the electronic subsystem, and how
the other terms in the phonon evolution equation (typically involving coupling with a thermal 
reservoir) try to bring this back to equilibrium. In the next section, we will, in fact, include
these `hot' phonon equations explicitly; instead of separate coupled equations, however, we will
treat the electron-phonon as a composite system, and use the couplings between the electrons, 
phonons, and with the contacts and the bath as processes driving the evolution of the composite 
system. 

The NEGF formalism also allows us to get away from this phonon-assisted tunneling limit to the
off-resonant limit. When the electronic levels lie far from resonance, their sidebands do not 
show up and we get instead the 
inelastic electronic tunneling spectrum (IETS) of the molecular system. For weak electron-phonon coupling (retaining 
leading order terms in $D_0$), the NEGF algebra simplifies considerably, so that the current 
partitions into an elastic component given by the usual Landauer formula, as well as an inelastic 
component that explicitly involves exclusion principle terms at different emission and absorption
energies. For applied voltages larger than the phonon frequency, these terms create additional
current transport channels through phonon emission, creating a slight increase in the current
that only shows up as peaks in the second derivative with voltage, generating the familiar IETS
spectrum. The physics resides entirely in the inelastic current -- one can use sophisticated 
electron structure methods to extract these peak positions from the phonon frequencies, as well
as the IETS peak heights from the computed deformation potentials. The results also show additional
subtleties near resonance that are experimentally observed. Specifically, near resonance the 
elastic current also picks up phonon signatures from the phonon contributions to $\Sigma$ residing 
in $G$, {\it{generating a dip rather than a peak}} that arises from the phonon sidebands described
above (the equations also generate a familiar polaronic shift of the main peak). The inclusion of
the equilibrium $n_B$ distribution gives phonon emission peaks that are stronger in strength than
absorption peak by Boltzmann ratios evaluated at the electronic temperature.  
 
While encouraging agreement between computed and experimental off-resonant behavior, specifically,
IETS spectra has been reported \cite{rpaulsson},
the resonant phonon sidebands and the scaling of their conductance peak heights with current
shows significantly more complex behavior arising from strong non-equilibrium electron-phonon 
correlations, which necessitates a Fock space approach. We will now introduce the joint electron-phonon 
Fock space and the experimental ramifications of electron-phonon correlation effects captured with this
treatment.  

\subsection{B. Fock space treatment of phonon sidebands}

\begin{figure}
\centerline{\epsfig{figure=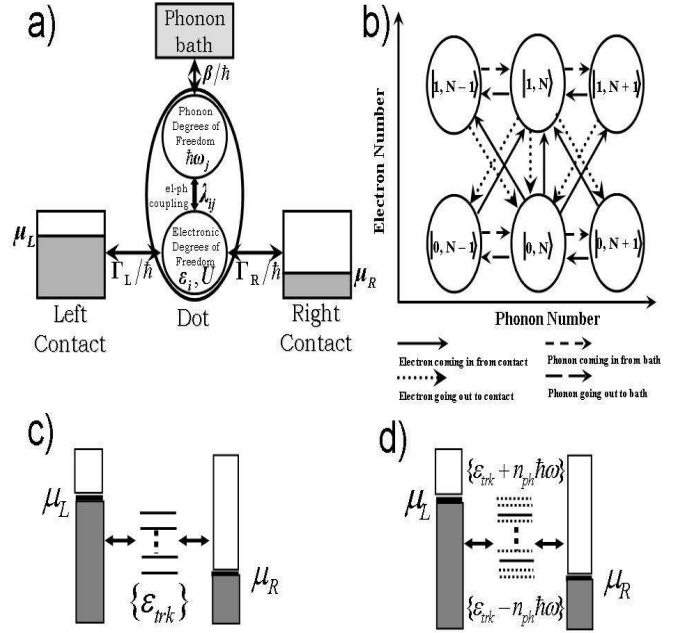,height=3.5in,width=3.5in}}
\caption{Fock space model to include electron-phonon interactions. a) The dot is electrically connected to the left 
(right) contact (with electron tunneling rates $\Gamma_{L,R}/\hbar$) and mechanically to the phonon bath (with a 
phonon escape rate of $\beta/\hbar$). The dot has electronic degrees of freedom $\epsilon_i$ and phonon degrees of 
freedom $\hbar\omega_j$ ($i, j=1, 2, 3, \ldots$) with coupling $\lambda_{ij}$. b) Transition between different states 
of a dot, having one electronic level and one phonon mode, due to coupling to the contact and the phonon bath 
($|0,N>$ and $|1,N>$ denotes the states of the system having $N$ phonons each but $0$ electron and $1$ electron 
respectively). c) Effective Coulomb blockaded channel in the absence of phonons. d) Inclusion of electron-phonon 
coupling adds phonon side bands to the existing transport channels.} 
\label{fig_7}
\end{figure}

We start from a model Hamiltonian for a quantum dot having onsite energies $\epsilon_i$, Coulomb interaction energy $U_{ii'}$, 
vibronic modes at energy $\hbar\omega_j$ and electron-phonon coupling $\lambda_{ij}$. The system is connected to 
electrical contacts with couplings $\Gamma_{1,2}$ and to a thermal bath with coupling $\beta$ 
(figure~\ref{fig_7}(a)). Electronic transport due to bias applied to the contacts as well as phonon emission and
absorption processes lead to transitions between many-body states $|e_{Ne}^i,k\rangle$ ($k$ phonons and {\it{i}}th 
electronic level in the $N_e$ electronic subspace) of the quantum dot (figure~\ref{fig_7}(b)). The rates of these 
transitions due to left (right) contact ($\mathcal{R}^{L(R)}_{|e_{Ne-1}^r,k\rangle\to|e_{Ne}^s,p\rangle}$) and 
the phonon bath ($\mathcal{R}^{ph}_{|e_{Ne-1}^r,k\rangle\to|e_{Ne}^s,p\rangle}$) are calculated by applying Fermi's 
Golden Rule~\cite{lutfe}. When an electron is added or removed via standard transport processes, it can also change 
the number of phonons, taking the dot from a state $|0,N\rangle$ to $|1,N \pm p\rangle$ or vice versa, where $p$ is 
the number of phonons emitted or absorbed. When the quantum dot absorbs or emits a phonon, the state changes 
from $|X,N\rangle$ to $|X,N \pm 1\rangle$, where $X=0$ or $1$. The consequence of phonons in transport is the 
addition of extra transport channels to the already existing Coulomb Blockade transport channels, as shown in 
Figures~\ref{fig_7}(c) and (d). 

The rate equations in the above Fock space (\ref{fig_7}) can be solved for a finite number of phonon emission or
absorption channels to extract the resulting current-voltage characteristic. Focusing on the intriguing experiments 
on Coulomb Blockaded nanotube quantum dots with prominent breathing modes \cite{leroy}, one sees multiple
intriguing features: (a) the absorption and emission sideband heights do not scale as simply as above. This is
because the number of phonons $N_{ph}$ itself changes with current, in addition to being driven far from equilibrium
in the suspended sections of the tube with small escape rate $\beta$. (b) The scaling of sidebands with phonon
population differs significantly from predictions of the analogous Tien-Gordon theory of photon-assisted 
tunneling (PAT) \cite{rtg}. This discrepancy arises because unlike the PAT experiments, the phonons are not 
coherent, and are partly correlated with the nanotube electrons \cite{lutfe}

\subsection{C. Non-equilibrium phonons and effective temperature} 
\begin{figure}
\centerline{\epsfig{figure=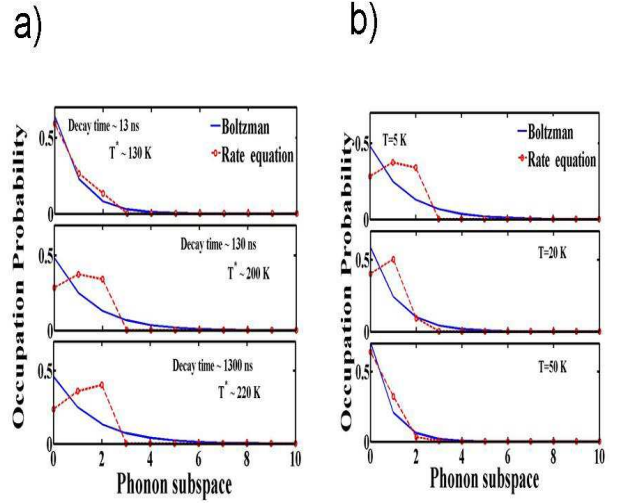,height=3.5in,width=3.5in}}
\caption{The phonon distribution $P^{phB}_k$ from the Boltzmann dist. a)({\it{solid line}}) with 
the effective temperature $T^{*}$ differs considerably from the phonon distribution $P^{ph}_k$ 
calculated directly from the rate equation ({\it{circles}}). b) The phonon distribution $P^{ph}_{k}$ approaches 
the Boltzmann distribution $P^{phB}_{k}$ (with proper effective temperature $T^{*}$ not shown here) at elevated surrounding temperature $T$.} 
\label{fig_8}
\end{figure}

A crucial component of the above picture is the strongly non-equilibrium distribution for the phonon population,
as their generation rate by the drive current (determined by $\Gamma_{L,R}$ and $\lambda$) exceeds their extraction 
rate $\beta$. From the solution to the rate equations giving us the joint electron-phonon occupation probabilities, 
we can calculate the probability distribution of the phonon by summing over the electronic subspaces: 
$P^{ph}_{k}=\sum_{s,Ne}P_{|e^s_{Ne},k\rangle}$. Defining an effective temperature $T^{*}$ 
corresponding to the non-equilibrium phonon occupation $N_{ph}=[e^{\hbar\omega/k_{B}T^{*}}-1]^{-1}$, we find 
the corresponding Boltzmann distribution: $P^{phB}_{k}=e^{-k\hbar\omega/k_{B}T^{*}}/Z$ of phonon subspace occupation probability, 
where $Z$ is the partition function ($T^*$ is identified by fitting the higher energy tails between the two
distributions). A comparison between $P^{ph}_{k}$ and $P^{phB}_{k}$ at 
different bias for different decay rates 
of the phonons reveals that they differ considerably as the phonon decay rate $\beta/\hbar$ decreases 
(figure~\ref{fig_8}(a)) at some applied bias. 
The deviation from a Bose-Einstein like shape suggests that the phonons in the suspended nanotubes are strongly 
non-equilibrium, so that {\it{temperature is not a well-defined quantity except to describe the higher energy 
tails}} (figure~\ref{fig_8}(b)). 

\subsection{D. Effect of the surrounding temperature: anomalous phonon population} 
\begin{figure}
\centerline{\epsfig{figure=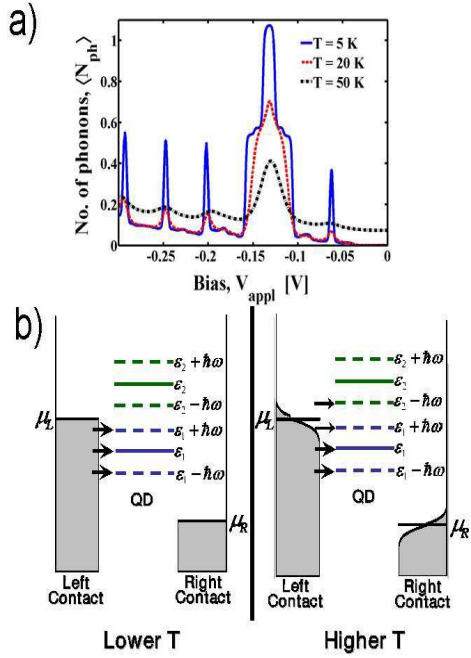,height=3.5in,width=3.5in}}
\caption{Anamolous temperature dependence of phonon distribution a) The number of phonons decreases at some bias points with increasing surrounding temperature $T$. (b) {\it{Bottom Left}}At lower surrounding temperature 
excitation of energy $\epsilon_1+\hbar\omega$ associated with resonant level $\epsilon_1$ falls 
inside the bias window between $\mu_L+k_BT_L$ and $\mu_R-k_BT_R$ and the excitation of energy $\epsilon_2-\hbar\omega$ 
associated with resonant level $\epsilon_2$ falls outside the bias window. {\it{Bottom right}}: 
At elevated surrounding temperature fermi functions of the contacts broadens around respective chemical potentials 
and the excitation of energy $\epsilon_2-\hbar\omega$ enters the bias window and lesser no. of electrons appear 
at energy $\epsilon_1+\hbar\omega$ in the left contact.} 
\label{fig_9}
\end{figure}

The non-equilibrium phonon population reveals a 
striking and somewhat counter-intuitive dependence on the substrate temperature, {\it{the phonon population 
decreasing with increasing temperature at certain bias voltages}} (Fig.~\ref{fig_9}(a)). 
This peculiar behaviour manifests itself as long as the surrounding temperature 
is smaller than the separation between the upper emission and lower absorption sidebands of two subsequent 
Coulomb Blockade peaks.
Beyond this temperature the phonon occupancy 
increases monotonically with temperature for all bias values, as expected.
The anomalous temperature dependence arises from a trade-off between 
phonon generation and recombination rates in the coupled dot-lead-bath system. The behavior 
is observed specifically at bias values corresponding to the onset of a new phonon absorption channel 
(Figure.~\ref{fig_9}(b), bottom left). 
With increasing temperature from $\sim$ 5 to 20K, the increasing tails of the contact Fermi functions 
redistribute the electrons from a phonon emission sideband $\epsilon_1 + \hbar\omega$ of a lower electronic peak 
to an absorption sideband $\epsilon_2 - \hbar\omega$ of a higher peak (Figure.~\ref{fig_9}(b) bottom right). 
Under this specific conspiracy of temperature and bias values, the number of electrons resonant with the phonon 
sidebands decrease with increasing temperature, decreasing the efficiency of the phonon-assisted transport. At higher temperatures between 20-50 K, the 
temperature is large enough to open new emission channels that eventually increases the phonon population 
at all bias voltages.

\section{Concluding Remarks}
For a wide variety of transport problems, a perturbative treatment of 
interactions coupled with a quantum kinetic
theory (such as NEGF) does an admirable job of explaining and predicting 
experimental features and providing
important physical insights. As objects scale towards nanodimensions, 
however, strong confinement and poor coupling
with the surroundings lead to increasing degrees of correlation, 
especially at lower temperatures. In this limit,
the Fock space approach (more generally, the many-body density matrix 
approach) naturally allows us to compute
transport signatures, provided we have a suitable means of extracting 
the various correlated many-body states and
the transition rates among them. Exact diagonalization provides one 
option, although this quickly becomes computationally intractable. 
Partial configuration interaction (CI) schemes may prove to be more 
practical. In
contrast with the NEGF-SCF limit where one could aim for quantitative 
and predictive accuracy with increasing
amounts of chemical sophistication, such details are hard to build into 
the Fock space approach and need to be
replaced by simpler, model problems, with parameters that could be 
benchmarked with more detailed models and
measurements. However, even these simple models (with a reasonable 
choice of parameters) show qualitatively new
physics that is experimentally observable, ranging from gate-modulated 
current levels, gate tunable excitation
spectra, scan-rate dependent NDR and hysteresis, and the breakdown of 
our common intuition based on equilibrium
phonon sideband scaling and the classical theories of photon-assisted 
tunneling.
While these two limits (quantum wire and quantum dot) are separately 
well understood, at least formally, the
intermediate coupling regime between the two becomes particularly 
challenging to model as there is no small `fine structure' parameter 
that would allow a convenient starting point for a perturbation 
expansion (e.g. a noninteracting wire or a fully interacting but 
isolated quantum dot). Significant progress is needed at formal, 
computational
and experimental levels in order to probe this regime, which bears the 
promise of completely novel physics of
non-equilibrium correlations as well as possible applications based on 
the interaction between conducting
detector elements in the quantum wire regime and non-conducting storage 
elements in the quantum dot regime. 

Acknowledgements: It is our pleasure to acknowledge Prof. Supriyo Datta for useful discussions.


\begin{thebibliography}{100}
\bibitem{tddft} In principle, a nonlocal time-dependent density functional theory (TDDFT) could capture these 
excitations even using an $N \times N$ one-electron potential matrix, provided the time-dependence
of the functional is rich enough to capture the additional excitations of the $2^N \times 2^N$ space
through its poles. In practice, however, this has not been done, even for the simplest system of two spin-degenerate levels.
\end{thebibliography}
\end{document}